\documentclass[twocolumn]{aa}
\usepackage{natbib,twoopt}
\usepackage{amsmath}

\usepackage[varg]{txfonts}
\usepackage{ulem}
\usepackage[breaklinks=true]{hyperref} 
\bibpunct{(}{)}{;}{a}{}{,}             
\makeatletter
  \newcommandtwoopt{\citeads}[3][][]{\href{http://adsabs.harvard.edu/abs/#3}%
    {\def\hyper@linkstart##1##2{}%
     \let\hyper@linkend\@empty\citealp[#1][#2]{#3}}}
  \newcommandtwoopt{\citepads}[3][][]{\href{http://adsabs.harvard.edu/abs/#3}%
    {\def\hyper@linkstart##1##2{}%
     \let\hyper@linkend\@empty\citep[#1][#2]{#3}}}
  \newcommandtwoopt{\citetads}[3][][]{\href{http://adsabs.harvard.edu/abs/#3}%
    {\def\hyper@linkstart##1##2{}%
     \let\hyper@linkend\@empty\citet[#1][#2]{#3}}}
  \newcommandtwoopt{\citeyearads}[3][][]%
    {\href{http://adsabs.harvard.edu/abs/#3}
    {\def\hyper@linkstart##1##2{}%
     \let\hyper@linkend\@empty\citeyear[#1][#2]{#3}}}
\makeatother

\usepackage{xcolor}

\begin{document}

\title{Quasi-periodic oscillation detected in $\gamma$-rays in blazar PKS 0346-27}
\titlerunning{ QPO in blazar}
   \authorrunning{R. Prince et al.}

\author{Raj Prince\inst{1} 
  \thanks{raj@cft.edu.pl}
     \and Anuvab Banerjee\inst{2}
     \and Ajay Sharma \inst{2}
     \and Avik Kumar das\inst{3}
     \and Alok C. Gupta\inst{4,5}
     \and Debanjan Bose\inst{6}}


\institute{ Center for Theoretical Physics, Polish Academy of Sciences, Al. Lotnik\'{o}w 32/46,
02-668 Warsaw, Poland
\and S. N. Bose National Centre for Basic Sciences, Block JD, Salt Lake, Kolkata 700106, India
\and Department of Physical Sciences, Indian Institute of Science Education and Research Mohali,
Knowledge City, Sector 81, SAS Nagar, Punjab 140306, India
\and Key Laboratory for Research in Galaxies and Cosmology, Shanghai Astronomical Observatory, Chinese Academy of Sciences, Shanghai 200030, China
\and Aryabhatta Research Institute of Observational Sciences (ARIES), Manora Peak, Nainital 263001, Uttarakhand, India
\and School of Astrophysics, Presidency University, 86/1 College Street, Kolkata 700073, West Bengal, India 
}
\date{}

\abstract {} {We present a variability study of the blazar PKS 0346-27 from December 2018 to January 2022 in its archival $\gamma$-ray observation by Fermi-LAT.} 
{We use the Lomb-Scargle periodogram  and the weighted wavelet transform methods in order to detect the presence of periodicity/quasi-periodicity and localize this feature in time and frequency space. The significance of the periodicity feature has been estimated using the Monte-Carlo simulation approach. We have also determined the global significance of the periodicity to test the robustness of our claim. To explore the most probable scenario, we modeled the light curve by both a straight jet and a curved jet model.} 
{We detect a periodicity feature of $\sim$ 100 days duration for the entire period of observation with a statistical significance of $3\sigma$, which amounts to a 99.7\% confidence level. The global significance of this feature is found to be 96.96\%. Based on the Akaike Information Criteria, the most probable explanation is that the observed emission is enhanced due to the helical motion of a blob within a curved jet.}
{The origin of this QPO is very likely a region of enhanced emission moving helically inside a curved jet. This work presents strong evidence for jet curvature
in the source and an independent (albeit a little serendipitous) procedure to estimate the curvature in blazar jets.}

\keywords{galaxies: active --
  BL Lacertae objects: individual: PKS 0346-27 -- BL Lacertae objects: general -- gamma rays: galaxies} 
\maketitle

\section{Introduction} \label{sec:intro}
Active galactic nuclei (AGN) are believed to derive their ultimate power from accretion onto a supermassive black hole (SMBH) with a mass in the range of 10$^{6} - \rm{10}^{10} \rm{M}_\odot$. 
Blazar variability can be sufficiently characterized as a red-noise process.
In the time series data, or light curves (LCs), quasi-periodic oscillations (QPOs) appear to be quite rare for AGNs \citep[see for a review][]{2018Galax...6....1G}.  There have been some strong claims of AGN QPOs in different bands of the EM spectrum with diverse periods \citep[e.g.,][and references therein]{2008Natur.455..369G,2009ApJ...690..216G,2018A&A...616L...6G,2019MNRAS.484.5785G,2009A&A...506L..17L,2013ApJ...776L..10L,2013MNRAS.436L.114K,2014MNRAS.445L..16A,2015MNRAS.449..467A,Sandrinelli2014,2016AJ....151...54S,2017A&A...600A.132S,2015Natur.518...74G,2015ApJ...813L..41A,2016ApJ...819L..19P,2019MNRAS.487.3990B,2021MNRAS.501...50S,2022MNRAS.513.5238R,2022Natur.609..265J}. However, many of the claimed QPOs, particularly those made earlier than 2008  are marginal detection, lasting only a few cycles, and the originally quoted statistical significance is probably overestimated \citep[see for a review][]{2014JApA...35..307G}. In the last $\sim$ 1.5 decades there have been very few strong claims of detection of QPOs on diverse timescales in blazars in different bands of the whole EM spectrum \citep[e.h][and references therein]{2009ApJ...690..216G,2019MNRAS.484.5785G,2009A&A...506L..17L,2013MNRAS.436L.114K,2015Natur.518...74G,2015ApJ...813L..41A,2018NatCo...9.4599Z,2019MNRAS.487.3990B,2021MNRAS.501...50S,2022MNRAS.513.5238R,2022Natur.609..265J}, and in other sub-classes of AGNs in X-ray bands \citep[e.g.,][and references therein]{2008Natur.455..369G,2014MNRAS.445L..16A,2015MNRAS.449..467A,2016ApJ...819L..19P,2018A&A...616L...6G}. \\ 
\\ 
Continuing analysis of blazars Fermi-LAT observations, there has been recent detection of QPOs in a few blazars on diverse timescales, e.g. $\sim$ 34.5 days QPO in PKS 2247–131  \citep{2018NatCo...9.4599Z}, $\sim$ 71 days QPO in B2 1520+31 \citep{2019MNRAS.484.5785G}, $\sim$ 47 days QPO in 3C 454.3 \citep{2021MNRAS.501...50S}, $\sim$ 3.6 days and $\sim$ 92 days QPOs in different segments of the LC of PKS 1510-089 \citep{2022MNRAS.510.3641R}. The first $\gamma$-ray QPO was detected in blazar PG 1553+113 by \citep{Ackermann2015, Tavani2018}. The QPO period was found to be $\sim$2.18 years with three number of cycle. The $\gamma$-ray QPOs of different time scales were also detected in many other blazars such as PKS 2155-304 (1.73 years) \citep{Sandrinelli2014, Zhang2017}, PKS 0426-380 (3.35 years), PKS 0301-243 (2.1 years) \citep{Zhang2017a,Zhang2017b}. A systematic search of QPO in the $\gamma$-ray LCs of blazars (FSRQ \& BL Lacs) is presented in \citep{Ren:2022bzt} where they detected a range of time scale of the order of months to years. A small sample of blazars is also explored by the \citep{Bhatta_2020} where the QPO in the gamma-ray LCs are found to be of the order of a few hundred days. The most interestingly $\gamma-$ray QPOs detected in all the blazars have periods in the range of a few tens of days to sometimes a year \citep{Ren:2022bzt}, and all blazars are FSRQs. It possibly gives evidence of the fundamental behavior of FSRQs which show $\gamma-$ray QPOs with these periods.\\

\begin{figure*} [h!]
    \centering
\includegraphics[width=14cm,height=6cm]{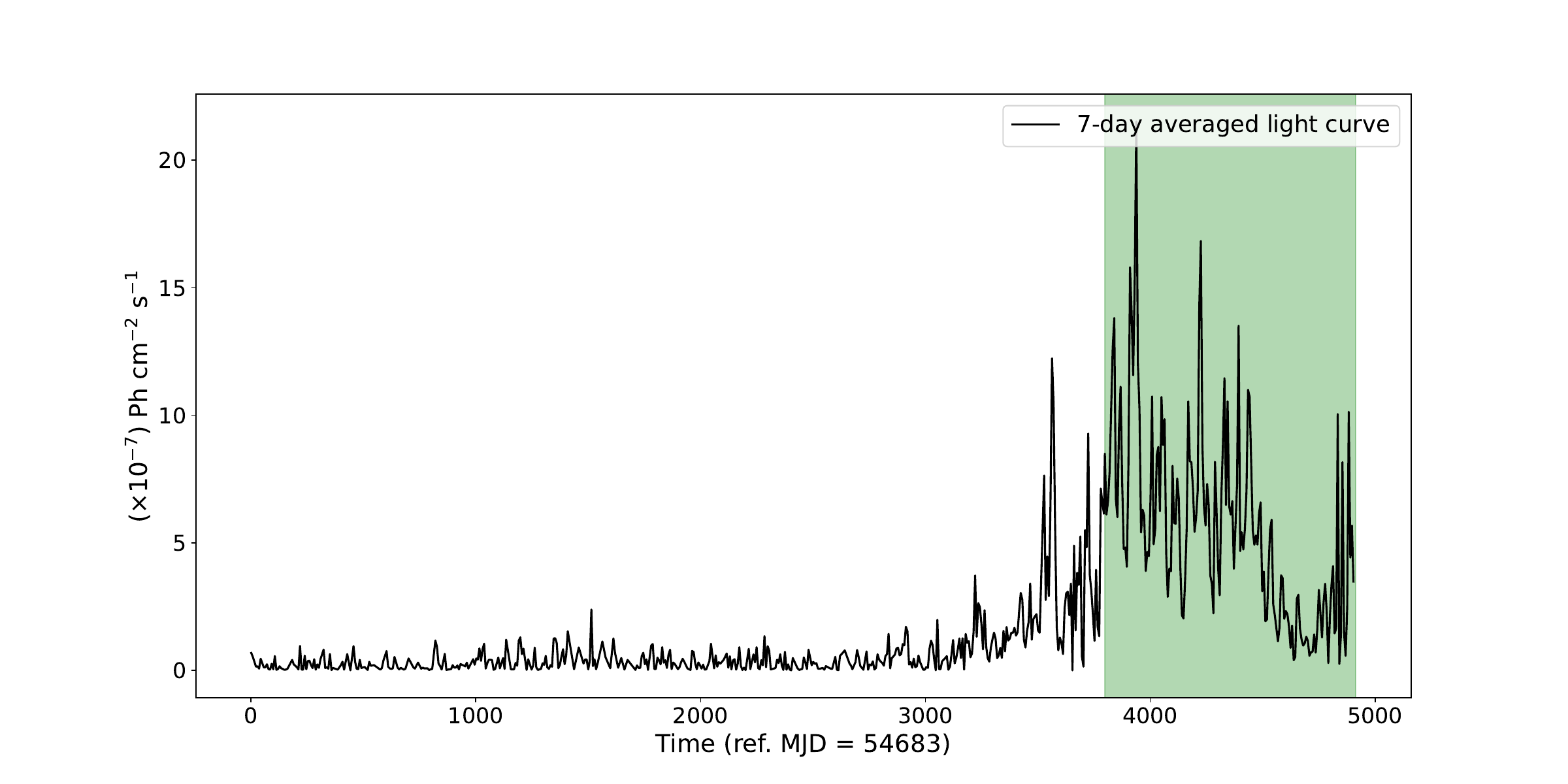}
  \includegraphics[width=14cm,height=6cm]{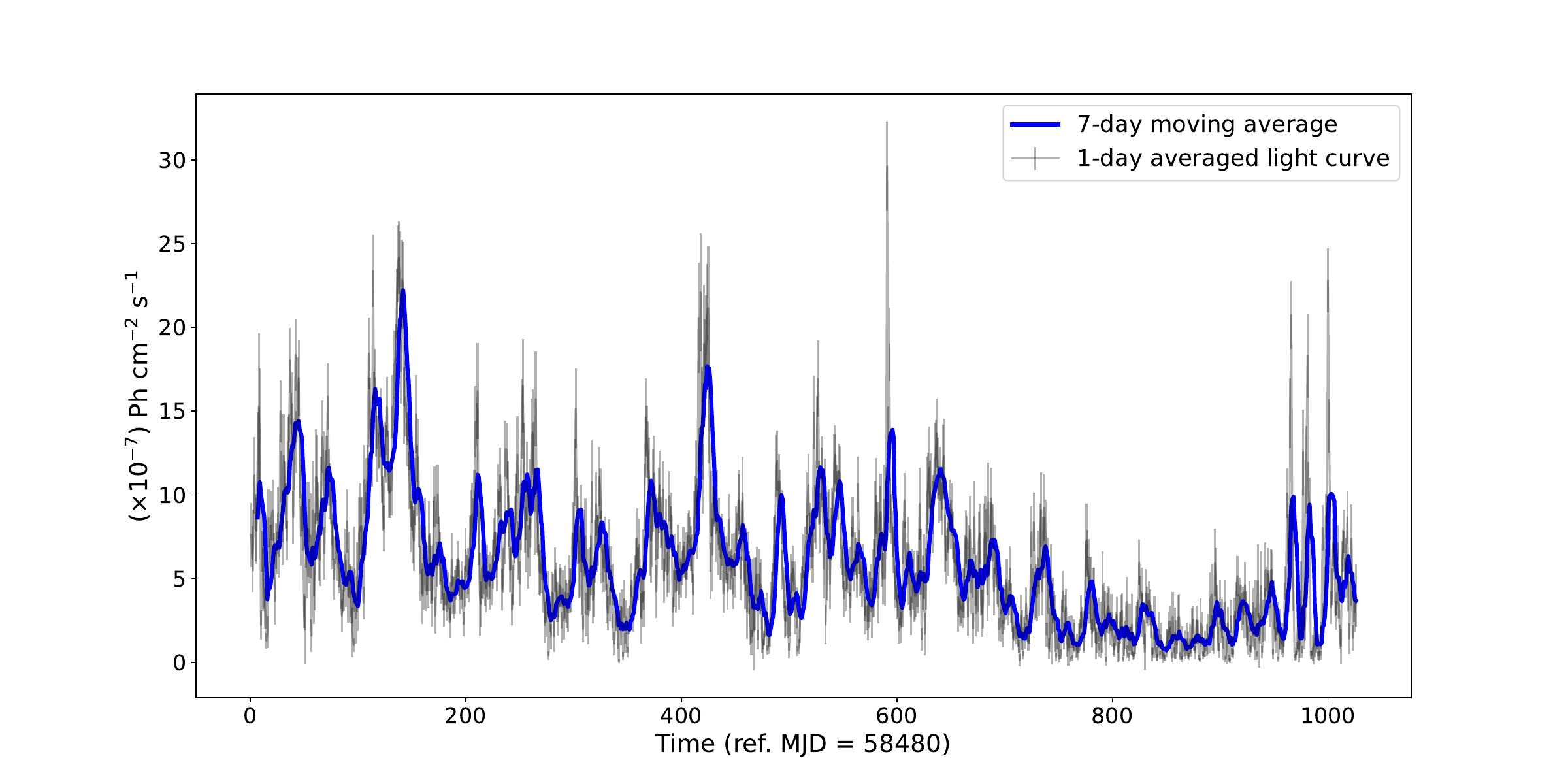}

    \caption{ The long-term $\gamma$-ray light curve for PKS 0346-27. For the majority of the time, the source was in the quiescent phase and has shown substantial flaring activity and a steady high flux state subsequent to December 2018. Upper panel: The 7-day binned light curve between August 5th, 2008 (MJD = 54683) and 19th Jan 2023 (MJD = 59598) and the high flux state is marked with a light green patch (MJD = 58480 to 59598) where QPO is searched. Lower panel: The 1-day binned Fermi-LAT light curve in 0.1-300 GeV for the full domain of observation between MJD = 58480 to 59598. We superimpose the 7-day moving average on top of the lightcurve, to make the periodic modulation explicit. From mere eye inspection, an oscillatory feature is observed in the light curve.}
    \label{combined}
\end{figure*}

\noindent
PKS 0346-27 is an FSRQ blazar at a redshift z = 0.991 \citep{1988ApJ...327..561W}.
The source has been detected in the different EM bands \citep{Sushanth_2023} and recently also discovered as a $\gamma-$ray emitter and listed in the Fermi-LAT Fourth Source Catalog (4FGL) \citep{2020ApJS..247...33A}. The mass of the SMBH at the center of PKS 0346-27 is estimated by multi-wavelength SED modeling in the low flux state of the source to be $\sim$ 2 $\times$ 10$^{8}$ M$_{\odot}$ \citep{2019A&A...627A.140A}. Multi-wavelength observation of this source has detected it in the high activity state in optical-NIR \citep{2018aNesciATel11269....1N}, as well as in UV and X-ray regions \citep{2018bNesciATel11455....1N}. Multi-waveband spectral investigation on quiescent and flaring phases using a one-zone leptonic model identified this source as intermediate frequency peaked (ISP) blazar \citep{2019A&A...627A.140A}. A detailed broadband SED modeling of the source is done in \citep{Sushanth_2023} where they have modeled the broadband SED with the combination of external Compton from the accretion disk and the broad-line region. \\
\\
The presence of QPOs in the LCs of AGNs is of great importance and can give strong support for the common nature of the accretion processes onto BHs ranging from a few solar masses up to the SMBHs present in AGNs \citep{2001A&A...374L..19A,2006ARA&A..44...49R,2015ApJ...798L...5Z}. The possible AGNs emission models which might explain QPOs in AGNs in different electromagnetic (EM) bands on diverse timescales have been discussed \citep[e.g.][and references therein]{2008Natur.455..369G,2009ApJ...690..216G,2009A&A...506L..17L,2016ApJ...819L..19P,2021MNRAS.501...50S,2022MNRAS.513.5238R,2022Natur.609..265J}. \\
\\
The draft is arranged as follows. In Section 2 of this Letter we briefly explained the Fermi-LAT $\gamma-$ray data and its analysis. In Section 3 we describe the various QPO analysis techniques and obtained results by those. The discussion and conclusion are provided in Section 4.

\section{Data and Reduction}\label{sec11}
We use the standard \texttt{Fermitools} package to extract and analyze the Fermi-LAT data in 0.1-300 GeV for the period between MJD 58480 to 59598. A 10$^\circ$ region of interest (ROI) was chosen around the source to extract the photon flux.
The user-specified constraints \texttt{`evclass=128' and `evtype=3'} corresponding to \texttt{gtselect} tool were applied to select rows from the input event data. The zenith angle cut of $90^{\circ}$ was applied to filter any plausible contamination from the Earth's limb. 
The recent 4FGL DR3 catalog was used to produce the source model.xml file and the source best-fit parameters were obtained using the maximum Likelihood.
The recent instrument response function \texttt{P8R3\_SOURCE\_V3} was applied for the purpose of the analysis. In order to account for the isotropic and the diffuse background emission, \texttt{iso\_P8R3\_SOURCE\_V3\_v1.txt} and \texttt{gll\_psc\_v31.fits} were employed respectively, available from Fermi Science Support Center (FSSC). 
We also identified the sources with low test statistics (TS<9) and were removed from further analysis. We fix all the other parameters of other sources within the ROI and optimize the source of interest parameters with a maximum Likelihood to obtain the flux in the defined time bin. We also estimate the TS value of each data point in the light curve and again apply the condition of TS<9 to remove the low photon statistics points.

\section{Light curve analysis and results}
 The $\gamma$-ray light curve of PKS 0346-27 in the energy range 0.1-300 GeV obtained from \textit{Fermi}-LAT is displayed in Figure~\ref{combined}(lower panel) shows a clear flux modulation decipherable from mere eye inspection during MJD 58480 to 59598. In the top panel, we present the full lightcurve starting from August 5th, 2008 (MJD = 54683), but it is observed that for most of the time, the source is in a quiescent phase and starts to show flux variability only subsequent to December 2018. The green-shaded region represents the domain of our analysis where the high flux state of the source, as well as variability features, are persistently observed for a duration of $>3$ years. In order to quantify such periodic modulation, we have adopted independent QPO analysis techniques, e.g. Lomb-Scargle periodogram (LSP), as well as Weighted Wavelet Z-transform (WWZ).


\subsection{Lomb-Scargle periodogram and testing the QPO significance}
 The Lomb-Scargle periodogram (LSP) method is one of the most widely used methods to identify the periodicities in time series \citep{Scargle1982ApJ...263..835S}. The advantage of the LSP over the standard discrete Fourier transform (DFT) method is that in the case of LSP, the data gaps and data collection  irregularities are accounted for by the least-square fitting of the sinusoidal waves of the form $X(t) = A\cos{\omega{t}} + B\sin{\omega{t}}$. Such a fitting procedure reduces the effect of the noise component and provides a correct measure of the detected periodicity \citep{zhang2017possible,zhang2017revisiting}. \\
\\ 
Red-noise type variability feature in temporal frequency is a hallmark of AGN or blazar sources. The periodogram is typically represented by a power spectral density (PSD) of the form $P(\nu) \sim A\nu^{-\beta}$, where $\nu$ represents the temporal frequency and $\beta > 0$ represents the spectral slope. Owing to such a power-law profile, the high amplitude features in the LSP detected at longer time scales (i.e. within the low-frequency domain) could appear to be a genuine periodicity feature \citep{1995A&A...300..707T,max2014method}. Owing to the presence of such a red-noise feature, rigorous estimation of the significance of periodogram peaks must be taken into consideration before concluding a peaked feature is a true QPO. We do the significance test by using the power spectral response (PSRESP; \citealt{Uttley2002}) method which is widely used in AGN and blazars \citep{Chatterjee_2008}. 
The red-noise PSDs of blazar lightcurves are typically well represented by a power-law or a bending power-law feature \citep{Vaughan2005Feb}. Therefore, we use a bending power law and a log-normal model respectively to fit the PSD and PDF of the original light curve. A total of 1000 light curves with the same PSD and PDF as that of the original light curve were simulated using the \texttt{DELightcurveSimulation}\footnote{\url{https://github.com/samconnolly/DELightcurveSimulation}} code. Using this procedure, we observe an evident peak at $\sim$ 100 days which is {\bf $\sim 3\sigma$} significant. 

\subsection{Weighted Wavelet Z-transform}
\begin{figure*}
    \centering
   \includegraphics[scale=0.19]{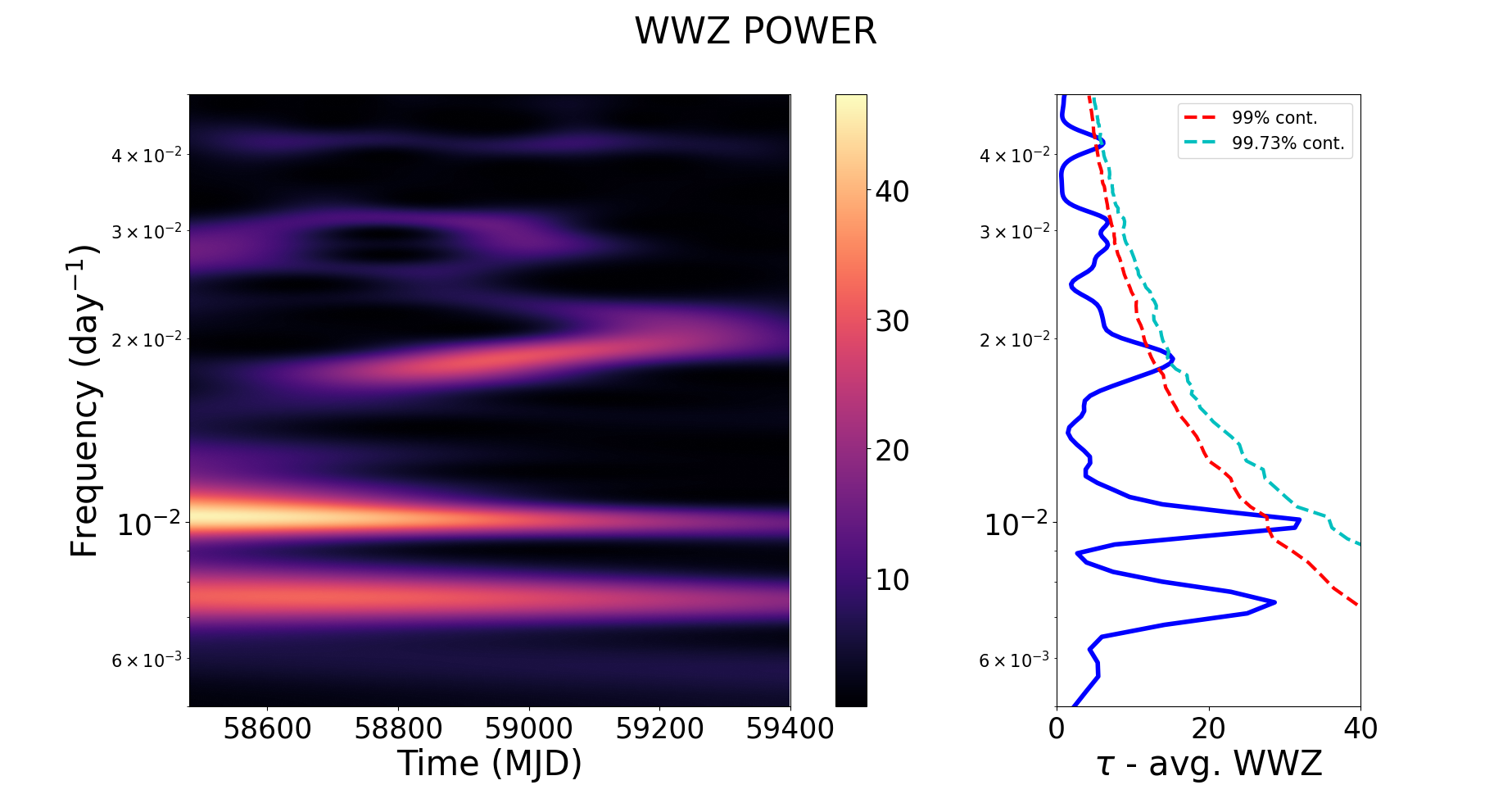}
   \includegraphics[scale=0.19]{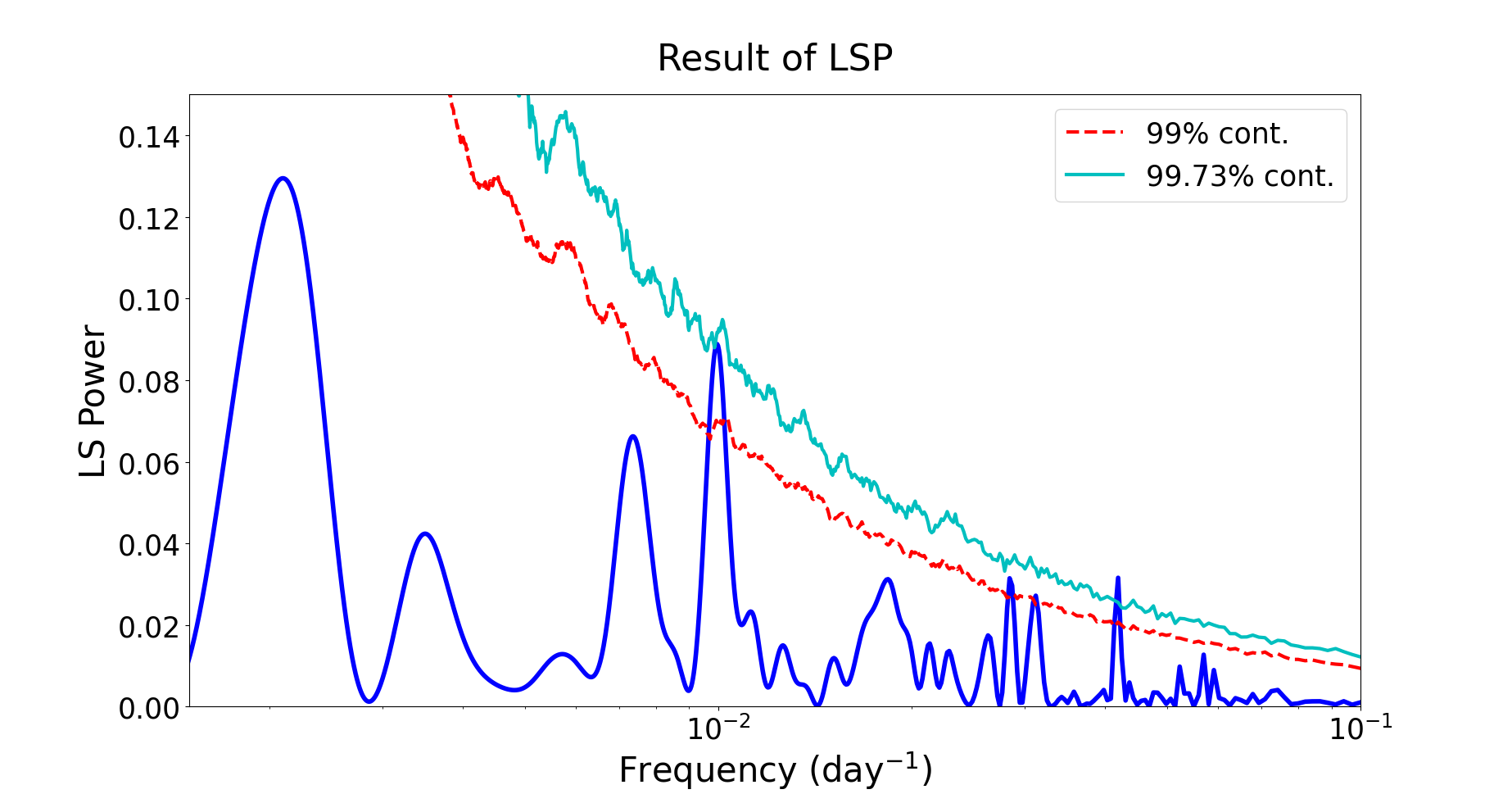}
    \caption{Plausible detection of the quasi-periodic oscillation feature corresponding to the lightcurve shown in the bottom panel in Figure~\ref {combined}. Right panel: We show the Lomb-Scargle periodogram with the statistical significance of the LSP peak, along with the average WWZ map. We observe the $0.01 \text{days}^{-1}$ peak to be $\sim 3\sigma$ significant. The average WWZ map shows a distinct peak at this frequency as well, strongly suggesting the presence of the QPO feature. (Left panel: We show the WWZ map along with the time-averaged power, where a distinct concentration of power within a narrow frequency band of around 100 days ($0.01 ~\text{days}^{-1}$) is observed. In both LSP and WWZ a 99.0\% and 99.73\% significance are estimated and shown in red and cyan color.}
    \label{lsp-wwz}
\end{figure*}

Another commonly used method of periodicity detection in blazar lightcurves is the wavelet transform method \citep{bhatta201372, bhatta2016detection, Das2022}. This approach attempts to determine the presence of any periodicity feature by fitting the data to sinusoidal. However, the localization of the waves in both time and frequency space is possible in this context to explore the evolution of QPO features with time \citep{Foster1996AJ....112.1709F}. It is a powerful tool to explore if such oscillations gradually develop, evolves in frequency space, and gradually dissipate over time.\\
\\
In brief, the Weighted Wavelet Z-transform (WWZ) method convolves a light curve with a time and frequency-dependent kernel and decomposes the data into time and frequency domains to create a WWZ map. We use the Morlet kernel \citep{GrossmannMorlet+2009+126+139} which has the functional form
\begin{equation}
    f[\omega(t-\tau)] = \exp[i\omega(t-\tau) - c\omega^2(t-\tau)^2].
\end{equation}
The corresponding WWZ map is
\begin{equation}
    W[\omega,\tau;x(t)] = \omega^{1/2}\int{x(t)f^{*}[\omega(t-\tau)]dt},
\end{equation}
where $f^{*}$ is the complex conjugate of the Morlet kernel $f$, and $\tau$ and $\omega$ are respectively the time and frequency shift. This kernel serves as a windowed discrete Fourier transform containing frequency dependent window of size $\exp{-c\omega^2(t-\tau)^2}$. The WWZ map has the advantage of
being able to detect statistically significant periodicities, as well as the time spans of their persistence. \\ 
 \\
 Our WWZ analysis results in a prominent peak at $\sim100$ days throughout the entire duration, as evident from the presence of significant power concentration within a narrow frequency window for the full domain of observation as shown in the right panel of Figure~\ref{lsp-wwz}. We also detected a broader peak in the WWZ map at $\sim$50 days but presents for a short period of time and it is well below the significance in the LSP method. 
 To claim the certainty of the QPO present in the light curve we have tested the significance of each peak. The significance of the peak was determined using the PSRESP method as described in the LSP methods, and the significance turns out to be $> 3\sigma$ (Figure-\ref{lsp-wwz}). 

\section{Discussion and Conclusion}
Several different physical models might explain the emergence of periodicity or quasi-periodicity in blazar light curves (LCs).
One plausible explanation could be a binary SMBH AGN system. According to this model, When the secondary BH pierces the primary BH's accretion disc during orbit, QPO may be detected \citep{2008Natur.452..851V}. This model is explicitly given for the blazar OJ 287 for which the mass of the binary SMBH system is much larger and the period is $\sim$ 12 years \citep{2008Natur.452..851V}. So, this model is unlikely to explain the detected QPO in the present work. Another possible explanation is the rotation of the accretion disk hot-spot or spiral shocks or some other non-axisymmetric phenomena around the innermost region of the accretion disk. This will primarily be manifested in the optical/X-ray domain and via external Compton scattering, periodicity in $\gamma$-ray photons could be observed. The central SMBH mass corresponding to this phenomenon happens to be \citep{2009ApJ...690..216G}
\begin{equation}
\frac{M}{M_\odot} = \frac{3.23\times10^4P}{(r^{3/2}+a)(1+z)},
\end{equation}
where $P$ is the orbital period in seconds, $z$ is the redshift of the object. The obtained ballpark for the central SMBH mass could be estimated in the case of a Schwarzschild BH (with $r = 6.0$ and $a = 0$), and for a maximal Kerr BH (with $r = 1.2$ and $a = 0.9982$) \citep{2009ApJ...690..216G}. In our case, considering the 100-day period, we get the central SMBH mass to be $9.48\times10^9~M_\odot$ in the Schwarzschild scenario and $6.01\times10^{10}~M_\odot$ in extreme Kerr scenario. The former estimate is quite large, and the later estimate corresponding to the Kerr scenario essentially exceeds all other SMBH mass estimates. Therefore, the variability feature to be directly reflective of some rotating axisymmetric phenomena is rather unlikely. \\
\\
 Magnetohydrodynamic origin of QPOs on a timescale of weeks to months driven by the kink instabilities in the jet spine has recently been proposed by \citep{dong2020MNRAS.494.1817D}. However, he also observed an anti-correlation between the optical polarisation degree (PD) and the light curve. However, at present, there is a lack of well-sampled optical PD data available to verify this scenario.
\par 
 Transient QPOs can also arise from a strong turbulent flow occurring behind a propagating shock or a standing shock in the jet \citep{marscher1992vob..conf...85M}. The dominant turbulent cell, which can exhibit enhanced Doppler boosting, introduces a QPO component to the observed light curve at the turnover period of the cell. In our case, this turnover period is $\sim$ 500 days (assuming Doppler factor = 10.0), suggesting the presence of a very large eddy necessary to explain the observed QPO. Furthermore, due to the stochastic nature of the cell, it is highly likely that the QPO would not persist for many cycles \citep{Wiita2011JApA...32..147W}.
\par
\begin{figure*}
    \centering
   \includegraphics[scale=0.20]{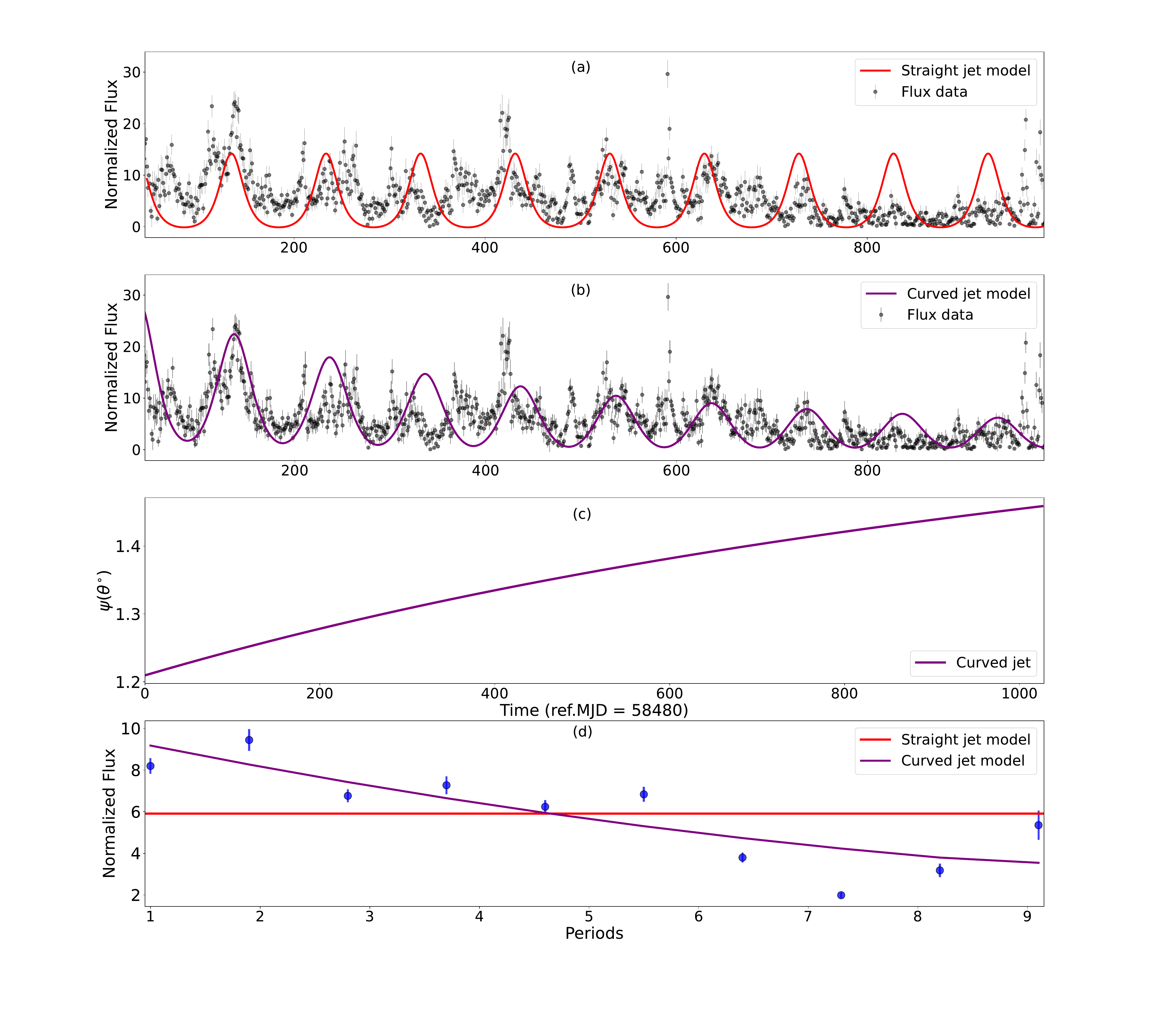}
    \caption{The gamma-ray light curve of PKS 0346-27 was modeled using two different jet scenarios: (a) a straight jet model (red) and (b) a curved jet model (purple) are shown with modulation of 100 days. The black data points represent the observed gamma-ray emission. (c) the viewing angle of the jet is shown as a function of time(t). (d) The average flux in each period was modeled using both the straight and curved jet models. The analysis of the Akaike information criteria strongly supports the curved jet scenario as the favored explanation for the observed gamma-ray emission within the given time domain. The straight jet model yielded an AIC value of 17.85, while the curved jet model, an AIC value is 10.15.}
    \label{jet-models}
\end{figure*}

\begin{figure} [h!]
    \centering
    \includegraphics[width=9.cm,trim=1.5cm 5.cm 0.6cm 4.cm,clip]{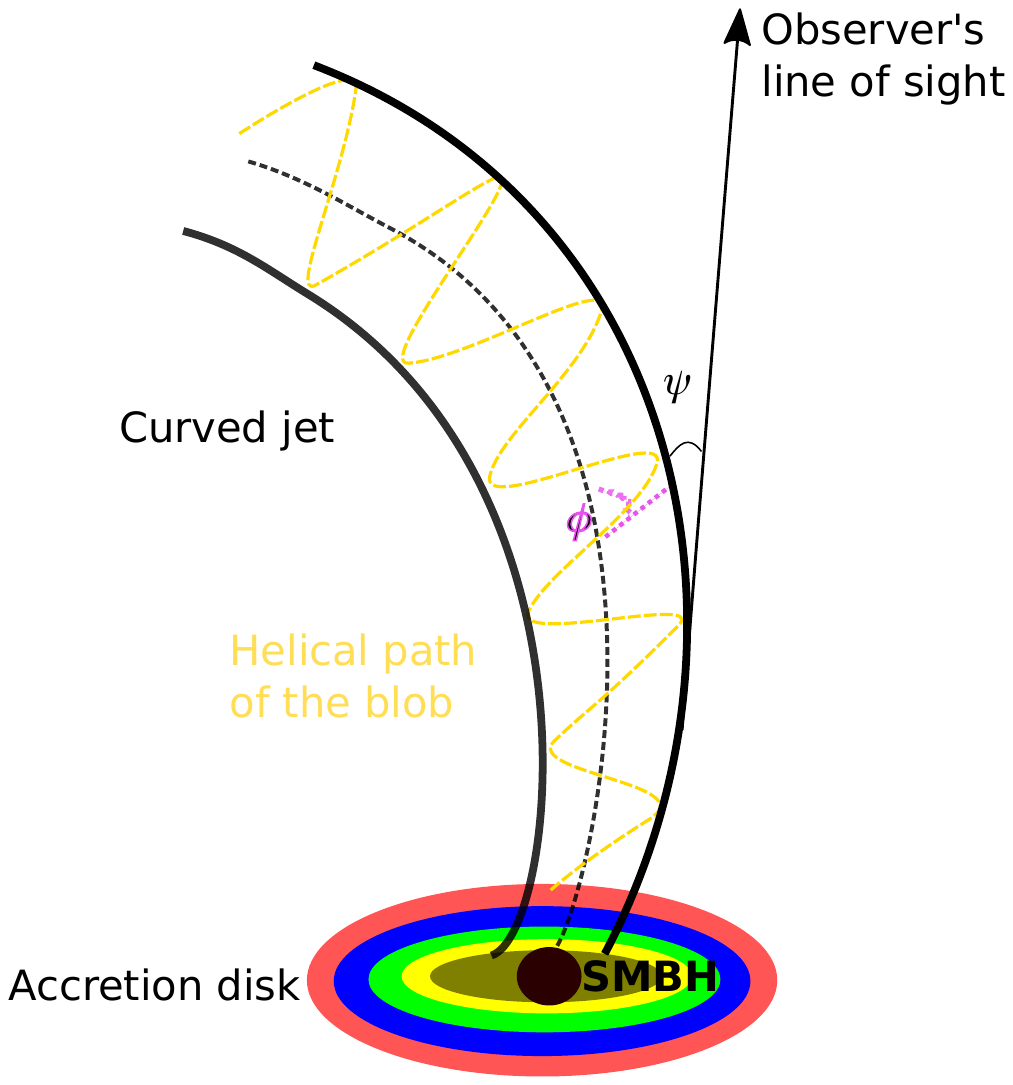}
    \caption{A possible curved-jet scenario present in the source is drawn here. The $\phi$ is the angle between the blob velocity vector and the jet axis. The $\psi$ is the viewing angle measured between the jet axis and the observer's line of sight. Note- Accretion disk is represented by the multi-color blackbody and the image is not to scale.}
    \label{curved}
\end{figure}

 
Since the blazar emission is jet dominated, it is quite likely that the variability signatures will have some connection with the jet emission characteristics. In the case of a precessing jet, quasi-periodic variability signatures could be clearly observed as a consequence of varying Lorentz factors along the line of sight of the observer. The jet precession could be induced by the presence of a secondary supermassive black hole (SMBH) in blazar creating a binary SMBH system \citep{Valtonen2008Natur.452..851V, Graham2015Natur.518...74G}, or if there is Lense-Thirring precession of the disc \citep{Fragile2009ApJ...693..771F} which would in turn influence the jet orientation. However, it has been suggested that such dynamical mechanisms would produce physical periods $\sim 1-2$ years \citep{Rieger2007Ap&SS.309..271R}, which is well above the periodicity we have observed in our case. \\
\\
Another jet-induced quasi-periodicity could originate from the motion of plasma blobs along the internal helical structure of the jet, as shown in Figure~\ref{curved}. The variation of the Doppler boosting factor as a consequence of the variation of the viewing angle of the plasma blob would lead to quasi-periodic variability features \citep{Mohan2015ApJ...805...91M}. Depending on the Doppler boosting factor and the viewing angle, the period of variability could vary from $\sim$ few days to a month time-scale. The periodicity that we have detected falls within this range. In this scenario, the blob can produce $\gamma$-ray emission via External Compton (EC) and Synchroton Self Compton (SSC) process (One-zone leptonic scenario). For a blob moving helically, the changing viewing angle of the blob with respect to our line of sight $cos\theta(t)$, is given by \citep{Zhou2018Nov},
\begin{equation} \label{eq:13}
\cos{\theta(t)} = \cos{\phi} \cos{\psi} + \sin{\phi} \sin{\psi} \cos({2 \pi t/ P_{obs}})
\end{equation}
Where $P_{obs}$ is the observed period and $\phi$ is the pitch angle defined between the blob velocity vector and the jet axis. The $\psi$ is the viewing angle or inclination angle measured between the observer's line of sight and the jet axis. As the observer, we see a boosted emission in gamma-ray with Doppler factor $\delta$, and hence the observed period in the blob frame can be translated as, $P_{obs}$ = (1 - $\beta$ cos$\psi$ cos$\phi$) $P^{'}$, where $P_{obs}$ is the observed period and the $P$ is the period in the blob frame. Given standard blazar parameters ($\psi$ \& $\phi$) we can estimate the period in the blob rest frame. For a FSRQ type blazar, the Lorentz factor is chosen as, $\Gamma$ = 15, $\phi$ = 2$^\circ$, and $\psi$ = 5$^\circ$. Using the expression, $\Gamma = 1 /\sqrt{1-\beta^2}$, $\beta$ is estimated to be 0.99777 and the QPO period in the blob rest frame is found to be, $P^{'}$ = 41.5 years. Along the helical path in the jet, the distance traveled by the blob in one cycle is given by $D_1$ = $c$ $\beta$ $P$ $cos \phi$ $\thickapprox$ 12.63 pc. 
The modified helical jet model introduces a significant departure from the conventional understanding of a straight jet, where the inclination angle of the jet's axis remains constant with respect to the line of sight. In this modified model, the blob exhibits helical motion within a curved jet, where the viewing angle $\psi \equiv \psi(t)$ {\citep{2021MNRAS.501...50S}} varies as a function of time. By incorporating the dependence of the viewing angle on time, expressed as $cos(\theta_{obs}(t))$, into the Doppler factor $\delta = 1/\Gamma(1 - \beta cos(\theta_{obs}))$, we obtain an expression for the observed emission ($F_{\nu} \propto \delta^3$) {\citep{2021MNRAS.501...50S}} as given
\begin{equation}
    {F_{\nu} \propto \frac{F_{\nu^{'}}^{'}}{\Gamma^3 (1 + sin \phi sin \psi)^3}\left( 1 - \frac{\beta cos\phi cos\psi}{1 + sin\phi sin\psi} cos\left(\frac{2\pi t}{P_{obs}}\right) \right)^{-3}}
\end{equation}

where $F_{\nu^{'}}^{'}$ is the rest-frame emission, $P_{obs}$ is the observed period. We conducted a comprehensive analysis to model the boosted emission in the observed frame using both a straight jet model (Figure \ref{jet-models}.a) and a curved jet model (Figure \ref{jet-models}.b). Additionally, we investigated the variation of the viewing angle over time (Figure \ref{jet-models}.c). Our analysis, based on the Akaike information criteria(AIC) (for the straight jet model, this value is found to be 17.85 and for the curved jet model, it is 10.15), favored that the curved jet model provides a more likely explanation for the observed gamma-ray emission within the given time domain (Figure \ref{jet-models}.d). Notably, including a multiplicative trend allowed us to account for changes in the Doppler factor, which could be attributed to spatial curvature in the jet. This curvature manifests as a time-dependent change in the viewing angle as the blob moves downstream. The hint of helical structure in the jet is also seen in the Very Long Baseline Interferometry (VLBI) in some blazars where the parsec-scale core appears to be misaligned with the kilo-parsec structure of the jet \citep{1993ApJ...411...89C}. 
In the absence of dense very long baseline interferometry (VLBI) measurements and comprehensive optical monitoring, our understanding of the viewing angle and jet Lorentz factor of this source is quite limited. Therefore, we do not attempt to constrain any of such jet-based models leading to $\gamma$-ray periodicity features.
As suggested by the Fermi science team the period modulation in the Fermi-LAT light curve can also come from the artifacts related to the telescope such as the precession period of the orbit of the Fermi spacecraft. To confirm that our result is not contaminated by this precession period, we have checked the Light curve of all the nearby sources within the 10$^\circ$ region of interest around the PKS 0346-27. It turns out that most of the sources within 10$^\circ$ are very faint and barely detected by Fermi-LAT. There are very few which have a reasonable light curve and we repeated our analysis on those light curves and did not find any evidence of periodic nature.

\begin{acknowledgements}
We thank the anonymous referee for their constructive criticism.
This research makes use of the publicly available data from \textit{Fermi}-LAT obtained from the FSSC data server and distributed by NASA Goddard Space Flight Center (GSFC). R. Prince is grateful for the support of the Polish Funding Agency National Science Centre, project 2017/26/A/ST9/-00756 (MAESTRO 9) and the European Research Council (ERC) under the European Union’s Horizon 2020 research and innovation program (grant agreement No. [951549]). A. Sharma is grateful to Prof. Sakuntala Chatterjee at S.N. Bose National Centre for Basic Sciences, for providing the necessary support to conduct this research.
ACG is partially supported by the Chinese Academy of Sciences (CAS) President’s International Fellowship Initiative (PIFI) (grant no. 2016VMB073). 
\end{acknowledgements}

\bibliographystyle{bibtex/aa.bst} 
\bibliography{references} 

\end{document}